\documentclass[review]{elsarticle}

\usepackage{subfigure}
\usepackage{amssymb}
\usepackage{mathtools}
\usepackage{multirow}
\usepackage{array}
\usepackage{booktabs}
\usepackage{tabularx}
\usepackage{longtable}
\usepackage{setspace}
\usepackage{graphicx}
\usepackage{diagbox}

\journal{Neural Networks}

\begin{document}

\begin{frontmatter}

\title{Correlating Subword Articulation with Lip Shapes for Embedding Aware Audio-Visual Speech Enhancement}

\author[USTC]{Hang Chen}
\author[USTC]{Jun Du\corref{mycorrespondingauthor}}
\cortext[mycorrespondingauthor]{Corresponding author}
\ead{jundu@ustc.edu.cn}
\author[USTC]{Yu Hu}
\author[USTC]{Li-Rong Dai}
\author[iflytek]{Bao-Cai Yin}
\author[GIT]{Chin-Hui Lee}

\address[USTC]{National Engineering Laboratory for Speech and Language
Information Processing, University of Science and Technology of China, Hefei, Anhui, China}
\address[iflytek]{iFlytek Research, iFlytek Co., Ltd., Hefei, Anhui, China}
\address[GIT]{School of Electrical and Computer Engineering, Georgia Institute of Technology, Atlanta, GA, USA}

\begin{abstract}
In this paper, we propose a visual embedding approach to improving embedding aware speech enhancement (EASE) by synchronizing visual lip frames at the phone and place of articulation levels. We first extract visual embedding from lip frames using a pre-trained phone or articulation place recognizer for visual-only EASE (VEASE). Next, we extract audio-visual embedding from noisy speech and lip videos in an information intersection manner, utilizing a complementarity of audio and visual features for multi-modal EASE (MEASE). Experiments on the TCD-TIMIT corpus corrupted by simulated additive noises show that our proposed subword based VEASE approach is more effective than conventional embedding at the word level. Moreover, visual embedding at the articulation place level, leveraging upon a high correlation between place of articulation and lip shapes, shows an even better performance than that at the phone level. Finally the proposed MEASE framework, incorporating both audio and visual embedding, yields significantly better speech quality and intelligibility than those obtained with the best visual-only and audio-only EASE systems.
\end{abstract}

\begin{keyword}
\texttt{Speech enhancement}\sep \texttt{audio-visual}\sep \texttt{representation learning} \sep \texttt{deep learning} \sep \texttt{universal attribute recognition}
\end{keyword}

\end{frontmatter}


\section{Introduction}\label{sec: sec_introduction}

 Background noises greatly reduce the quality and intelligibility of the speech signal, limiting the performance of speech-related applications in real-world conditions (e.g. automatic speech recognition, dialogue system and hearing aid, etc.). The goal of speech enhancement \cite{loizou2013speech} is to generate enhanced speech with better speech quality and clarity by suppressing background noise components in noisy speech.

 Conventional speech enhancement approaches, such as spectral subtraction \cite{boll1979suppression}, Wiener filtering \cite{lim1978all}, minimum mean squared error (MMSE) estimation \cite{ephraim1985speech}, and the optimally-modified log-spectral amplitude (OM-LSA) speech estimator \cite{cohen2001speech,cohen2003noise}, have been extensively studied in the past. Recently, deep learning technologies have been successfully used for speech enhancement \cite{6932438,6639038,8369155}.

 Human auditory system can track a single target voice source in extremely noisy acoustic environment like a cocktail party, as known as the cocktail party effect \cite{cherry1953some}. This fascinating nature motivates us to utilize the discovery that humans perceive speech when designing speech enhancement systems. McGurk effect \cite{mcgurk1976hearing} suggests a strong influence of vision in human speech perception. More researches \cite{bernstein1996speech, macleod1987quantifying,massaro2014speech,rosenblum2008speech} have shown visual cues such as facial/lip movements can supplement acoustic information of the corresponding speaker, helping speech perception, especially in noisy environments. Inspired by the above discoveries, the speech enhancement method utilizing both audio and visual signals, known as audio-visual speech enhancement (AVSE), has been developed.

 The AVSE methods can be traced back to \cite{girin1995noisy} and following work, e.g. \cite{girin2001audio,fisher2001learning,deligne2002audio,goecke2002noisy,hershey2002audio,abdelaziz2013twin}. And recently numerous studies have attempted to build deep neural network-based AVSE models. \cite{gabbay2018seeing} employed a video-to-speech method to construct T-F masks for speech enhancement. An encoder-decoder architecture was used in \cite{gabbay2018visual, hou2018audio}. These methods were merely demonstrated under constrained conditions (e.g. the utterances consisted of a fixed set of phrases, or a small number of known speakers). \cite{Afouras2018} proposed a deep AVSE network consisting of the magnitude and phase sub-networks, which enhanced magnitude and phase, respectively. \cite{Ephrat2018looking} designed a model that conditioned on the facial embedding of the source speaker and outputted the complex mask. \cite{8969244} proposed a time-domain AVSE framework based on Conv-tasnet \cite{luo2019conv}. These methods all performed well in the situations of unknown speakers and unknown noise types.

 We briefly discuss the above-mentioned AVSE methods from following two perspectives: visual embedding and audio-visual fusion method. Regrading to the visual embedding, \cite{hou2018audio,gabbay2018seeing,gabbay2018visual} made use of the image sequences of the lip region. For discarding irrelevant variations between images, such as illumination, \cite{Ephrat2018looking} proposed using the face embedding obtained from a pre-trained face recognizer and confirmed through ablation experiments that the lip area played the most important role for enhancement performance in the face area. Moreover, \cite{Afouras2018, 8969244} chose lip embedding via the middle layer output in a pre-trained isolated word recognition model.

 In recent work, \cite{wu2019time} adopted the phone as the classification target instead of isolated word and provided a more useful visual embedding for speech separation. In the term of audio-visual fusion method, most AVSE methods focus on audio-visual fusion that happens at the middle layer of the enhancement network in the fashion of channel-wise concatenation.

 We can get some inspirations from these pioneering works. A useful visual embedding should contain as much acoustic information in the video as possible. But the acoustic information in video is very limited, and there is also other information redundancy. In the current classification-based embedding extracting framework, we can yield a more robust and generalized visual embedding by reducing the information redundancy and increasing the correlation between the classification target and the visual acoustic information. Cutting out the lip area is helpful for reducing the redundancy. While for the other one, finding a classification target that is more relevant to lip movements is informative.

 The superset of speech information called speech attributes include a series of fundamental speech sounds with their linguistic interpretations, speaker characteristics, and emotional state etc \cite{lee2007overview}. In contrast to phone models, a smaller number of universal attribute units are needed for a complete characterization of any spoken language \cite{li2006vector}. The set of universal speech attributes used in related works mainly consist of place and manner of articulation \cite{li2005study, siniscalchi2013universal, li2016improving}. We propose that a higher correlation between the speech attribute and the visual acoustic information can provide a more useful supervisory signal in the stage of visual embedding extractor training.

 One commonly accepted consensus in multimodal learning is that the data of each mode obeys an independent distribution conditioned on the ground truth label \cite{blum1998combining,dasgupta2002pac,leskes2005value,lewis1998naive}. Each mode captures features related to ground truth tags from different aspects, so the information extracted (labels excluded) is not necessarily related to the other. This shows that the ground truth can be seen as ``information intersection" between all modes \cite{sun2020tcgm}, i.e., the amount of agreement shared by all the modalities. Specifically, in AVSE, there is a mismatch between the information intersection and the ground truth label. The intersection of audio modal (noisy speech) and video modal (lip video) is not clean speech which is the ground truth.

 In this work, we extend the previous AVSE framework to the embedding aware speech enhancement (EASE) framework. The conventional AVSE methods are regarded as special EASE methods, which only utilize visual embedding extracted from lip frames, as known as visual-only EASE (VEASE) methods. In EASE framework, we propose a VEASE model using a novel visual embedding, which is the middle layer output in a pre-trained universal place recognizer. We have the same dataset in the stages of the embedding extractor training and the enhancement network training. A more effective visual embedding is obtained by utilizing a high correlation between the designed classification target, i.e., the articulation place, and the visual acoustic information rather than additional video data. Moreover, we present a novel multimodal EASE (MEASE) model using multimodal embedding instead of unimodal embedding. The visual embedding extractor in the VEASE model evolves into audio-visual embedding extractor in the MEASE model. The enhancement network takes not only the noisy speech but also the fused audio-visual embedding as inputs, and outputs a ideal ratio mask. The fusion of audio and visual embeddings occurs in the stage of embedding extractor training and is supervised by their information intersection at the articulation place label level.

 The main contributions of this paper are:
\begin{enumerate}[(1)]
\item We explore the effectiveness of different visual embeddings pre-trained for various classification targets on enhancement performance. A novel classification target, i.e., the articulation place, is proposed for training visual embedding extractor. The visual embedding utilizing a high correlation between the articulation place and the acoustic information in video achieves the better enhancement performance with no additional data used.
\item We verify the complementarity between audio and visual embeddings lies in different SNR levels, as well as different articulation places by ablation experiments. And based on the information intersection, we adopt a novel fusion method integrating visual and audio embeddings in the proposed MEASE model, which achieves better performance in all SNR levels and all articulation places.
\item We design experiments to study the effect of the stage when audio-visual fusion occurs on the quality and intelligibility of enhanced speech. And we observe that the early fusion of audio and visual embeddings achieves the better enhancement performance.
\end{enumerate}

 Concurrently and independently from us, a number of groups have proposed various methods from above two perspectives for AVSE. \cite{wang2020robust} observed serious performance degradations when these AVSE methods were applied with a medium or high SNR \footnote{Performance degradation in \cite{wang2020robust} may result from the changes in the network structure, but we have indeed observed reduction in improvements from our results of comparative experiment, as will be discussed in Section \ref{sec: sec_MEASE_result}.} and proposed a late fusion-based approach to safely combine visual knowledge in speech enhancement. This is the opposite of our work. \cite{iuzzolino2020av} proposed a new mechanism for audio-visual fusion. In this research, the fusion block was adaptable to any middle layers of the enhancement network. This kind of multiple fusion in the enhancement network was better than the standard single channel-wise concatenation. However, their work differs ours in that audio-visual integration still occurs in the middle of the enhancement network.

 The rest of the paper is structured as follows. In Section \ref{sec: sec_VEASE_main} we describe the proposed VEASE method. The proposed MEASE method is presented in Section \ref{sec: sec_MEASE_main}. Section \ref{sec: sec_exp} has experimental setup including dataset, audio and video preprocessing as well as compares experimental results. Finally, we conclude this work and discuss future research directions in Section \ref{sec: sec_conclusion}.

\section{VEASE Model Utilizing Articulation Place Label}\label{sec: sec_VEASE_main}

 In this section, we elaborate our proposed VEASE model, including two aspects, i.e., architecture and training process. The visual embedding extractor is an important part of the VEASE model, which takes a sequence of lip frames as input and outputs a compact vector for every lip frame, known as visual embedding. The VEASE model takes both noisy log-power spectra (LPS) features and visual embeddings as inputs, and outputs ideal ratio mask. The details of the visual embedding extractor and the VEASE model are elaborated in the following.

\subsection{Architecture of Visual Embedding Extractor}\label{sec: sec_VEE_arc}

\begin{figure}[htb]
\centering
\includegraphics[width=\linewidth]{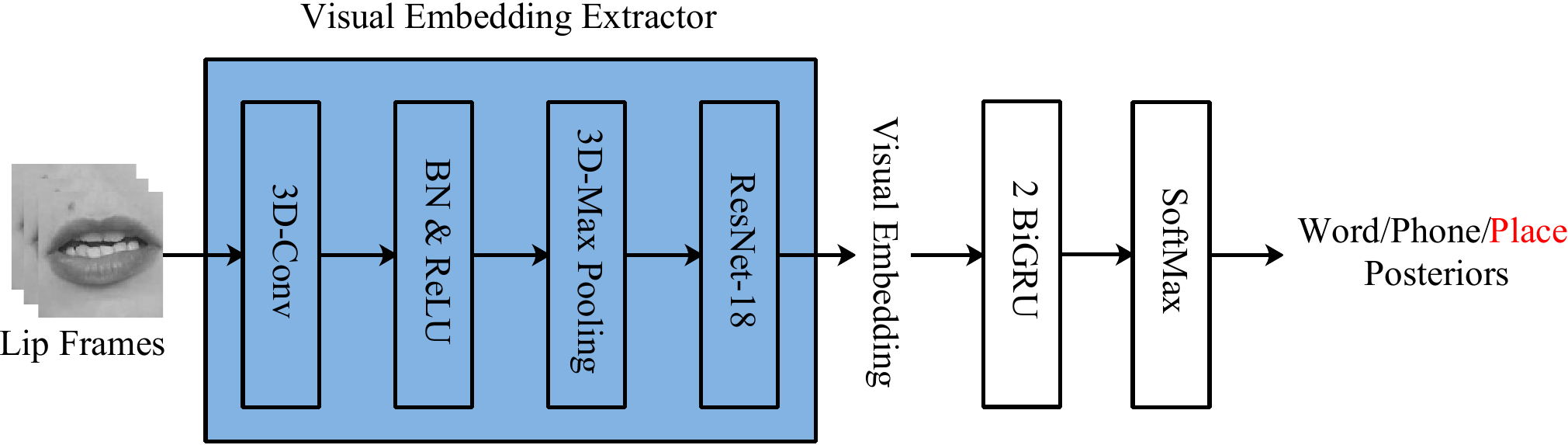}
\caption{Illustration of a visual embedding extractor (in color blue for ease of cross-referencing in Figure \ref{fig: fig_VEE}, Figure \ref{fig: fig_VEASE}, Figure \ref{fig: fig_MEASE} and Figure \ref{fig: fig_fuse_stage}). For every lip frame, the extractor outputs a compact vector. We train visual embedding extractor by using 3 different classification labels, i.e., word, phone and place.}
\label{fig: fig_VEE}
\end{figure}

 The visual embedding extractor $f_{\rm V}$ has a similar structure to  \cite{Stafylakis2017,petridis2018end}, which is also used in previous AVSE studies \cite{Afouras2018,8969244}. The extractor consists of a spatiotemporal convolution followed by an $18$-layer ResNet \cite{he2016deep} which is the identity mapping version \cite{he2016identity}, as shown in Figure \ref{fig: fig_VEE}. A spatiotemporal convolution consists of a convolution layer with $64$ 3D-kernels of $5\times7\times7$ (time/width/height), a batch normalization, a ReLU activation and a spatiotemporal max-pooling layer.

 For a sequence of lip frames $V=\{V^t\in \mathbb{R}^{H \times W}; t=0, 1, \dots, T_{\rm V}-1\}$, the feature maps is extracted by the spatiotemporal convolution. Then, the feature maps are passed through the $18$-layer ResNet. The spatial dimensionality shrinks progressively in the ResNet until output becomes a $L_{\rm V}$-dimensional vector per time step, known as the visual embedding $E_{\rm V}$:
\begin{equation}
\begin{split}
E_{\rm V} &= \{E_{\rm V}^t \in \mathbb{R}^{L_{\rm V}}; t=0, 1, \dots, T_{\rm V}-1\} = f_{\rm V}(V)\\
     &= \operatorname{ResNet-18}_{\rm V}(\operatorname{MaxPooling_{3D}}(\operatorname{BN}(\operatorname{ReLU}(\operatorname{Conv_{3D}}(V)))))
\end{split}
\end{equation}
where $T_{\rm V}$, $H$ and $W$ denote the number and the size of lip frames, respectively. In this study, we use $L_{\rm V}=256$, $H=98$ and $W=98$ by default.

 The visual embedding extractor is trained with a classification backend $f'_{\rm C}$ in the right side of Figure \ref{fig: fig_VEE}, consisting of a $2$-layer BiGRU, a fully connected layer followed by a SoftMax activation. The output of the $E_{\rm V}$ is fed to $f'_{\rm C}$ and the posterior probability of each class representing each segment of lip frames $P_{\rm word}$ is calculated as follows:
\begin{equation}
P_{\rm class} = f'_{\rm C}(E_{\rm V}) = \operatorname{SoftMax}(\operatorname{Mean}(\operatorname{FC}(\operatorname{BiGRU}(\operatorname{BiGRU}(E_{\rm V}))))),
\end{equation}
where the class can be labeled as word, phone or place of articulation.

\subsection{Word Based Visual Embedding Extraction}\label{sec: sec_VEE_word}

 Conventional AVSE techniques \cite{Afouras2018, 8969244} often obtain the visual embedding extractor discussed earlier based on an isolated word classification task by using a lip reading dataset, such as the Lip Reading in the Wild (LRW).
 
 We build our baseline model, denoted as VEASE-word using the LRW corpus consisting of up to $1000$ audio-visual speech segments extracted from BBC TV broadcasts (News, Talk Shows, etc.), totaling around $170$ hours. There are $500$ target words and more than $200$ speakers. The LRW dataset provides a word-level label for each audio-visual speech segment, i.e., the real distribution of word $P_{\rm word}^{\rm truth}$. We calculate the cross entropy (CE) loss $\mathcal{L}_{\rm CE}$ between $P_{\rm word}^{\rm truth}$ and $P_{\rm word}$:

\begin{equation}
\mathcal{L}_{\rm CE} = \operatorname{CE}(P_{\rm word}^{\rm truth} \| P_{\rm word}) = -\sum{P_{\rm word}^{\rm truth}\log P_{\rm word}}.
\label{eq: eq_word_ce}
\end{equation}

 The objective function, $\mathcal{L}_{\rm CE}$, is minimized by using Adam optimizer \cite{kingma2014adam} for $100$ epochs and the mini-batch size is set to $64$. The initial learning rate is set to $0.0003$ and is decreased on log scale after $30$ epochs. Data augmentation is performed during training, by applying random cropping ($\pm5$ pixels) and horizontal flips, which is the same across all lip frames of a sequence. The best model is selected by the highest frame-level classification accuracy.

\subsection{Phone Based Visual Embedding Extraction}\label{sec: sec_VEE_phone}

 The isolated word classification task usually requires a word-level dataset which is not easy to collect in a large scale effort. To alleviate this problem, we propose that the same data is used during training visual embedding extractor and enhancement network with different labels. Under the guidance of results in \cite{wu2019time}, we choose context-independent (CI) phones consisting of $39$ units from CMU dictionary as classification labels, denoted as VEASE-phone.

 $E_{\rm V}$ is fed to a classification backend $f_{\rm C}$ which has a same structure as $f'_{\rm C}$ and outputs the posterior probability of each CI-phone for each specific time frame $P_{\rm phone} = f_{\rm C}(E_{\rm V}) = \operatorname{SoftMax}(\operatorname{FC}(\operatorname{BiGRU}(\operatorname{BiGRU}(E_{\rm V}))))$.

 The TCD-TIMIT dataset is a high quality audio-visual speech corpus labeled at both the phonetic and the word level. We can directly get the frame-level real distribution of CI-phone $P_{\rm phone}^{\rm truth}$. The calculation of $\mathcal{L}_{\rm CE}$ between $P_{\rm phone}^{\rm truth}$ and $P_{\rm phone}$ is similar to Equation (\ref{eq: eq_word_ce}):
\begin{equation}
\mathcal{L}_{\rm CE} = \operatorname{CE}(P_{\rm phone}^{\rm truth} \| P_{\rm phone}) = -\sum{P_{\rm phone}^{\rm truth}\log P_{\rm phone}}
\label{eq: eq_phone_ce}
\end{equation}
We use the same optimization process as in Section \ref{sec: sec_VEE_word} to minimize $\mathcal{L}_{\rm CE}$.

\subsection{Articulation Place Based Visual Embedding Extraction}\label{sec: sec_VEE_place}

\begin{figure}[htb]
\centering
\includegraphics[width=\linewidth]{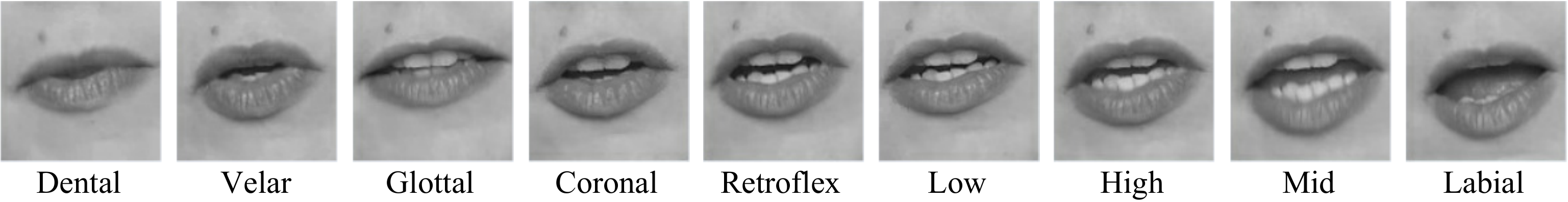}
\caption{9 lip shapes corresponding to utterance segments representing 9 articulation positions: all lip shapes come from a single speaker starting with the lip closed. The lip shape changes greatly in High, Mid and Labial than Dental, Velar and Glottal.}
\label{fig: fig_lip2place}
\end{figure}

 As discussed earlier, we believe there is a high correlation between speech attributes and visual acoustic information. In order to verify our idea, we check the lip shapes belonging to different places and manners of articulation. We find that the influences of various articulation places on the change of lip shape are different, i.e., the lip shape changes greatly in some utterance segments belonging to specific articulation place. An example is presented in Figure \ref{fig: fig_lip2place}. In contrast, we do not observe similar changes in the term of articulation manner. Consequently we propose to train visual embedding extractor with the articulation place label in this study, denoted as VEASE-place. We adopt 10 units as in \cite{siniscalchi2009study,lee2013information} for articulation place set.

 Compared with the phone, the category granularity of articulation place is coarser. Thus, the classification model can achieve comparable performance with lower complexity. And the articulation place has fewer categories, which reduces the labeling costs. Moreover, the articulation place label is believe to be more language-independent than phones, which allows various languages to appear in training and testing.

 The same classification backend $f_{\rm C}$ takes $E_{\rm V}$ as input and outputs the posterior probability of each articulation place class for each specific time frame $P_{\rm place}$.

\begin{table}[htb]
\centering
\caption{The mapping between articulation place classes and CI-phones as in \cite{siniscalchi2009study}.}
\label{tab: tab_phone2place}
\footnotesize
\setlength{\tabcolsep}{20pt}{
\begin{tabular}[b]{|l|l|}
\toprule
\textbf{Articulation place classes}&\textbf{CI-phones}\\
\midrule
Coronal&d, l, n, s, t, z\\
High&ch, ih, iy, jh, sh, uh, uw, y\\
Dental&dh, th\\
Glottal&hh\\
Labial&b, f, m, p, v, w\\
Low&aa, ae, aw, ay, oy\\
Mid&ah, eh, ey, ow\\
Retroflex&er, r\\
Velar&g, k, ng\\
Silence&sil\\
\bottomrule
\end{tabular}
}
\end{table}

 $P_{\rm phone}^{\rm truth}$ is mapped into the frame-level real distribution of articulation place $P_{\rm place}^{\rm truth}$ by using Table \ref{tab: tab_phone2place}. $\mathcal{L}_{\rm CE}$ between $P_{\rm place}^{\rm truth}$ and $P_{\rm place}$ is calculated as follows:
\begin{equation}
\mathcal{L}_{\rm CE} = \operatorname{CE}(P_{\rm place}^{\rm truth} \| P_{\rm place}) = -\sum{P_{\rm place}^{\rm truth}\log P_{\rm place}}
\label{eq: eq_place_ce}
\end{equation}
The optimization process to minimize $\mathcal{L}_{\rm CE}$ is same as in that in Section \ref{sec: sec_VEE_word}.

\subsection{VEASE Model}\label{sec: sec_VEASE}

\begin{figure}[htb]
\centering
\includegraphics[width=0.6\linewidth]{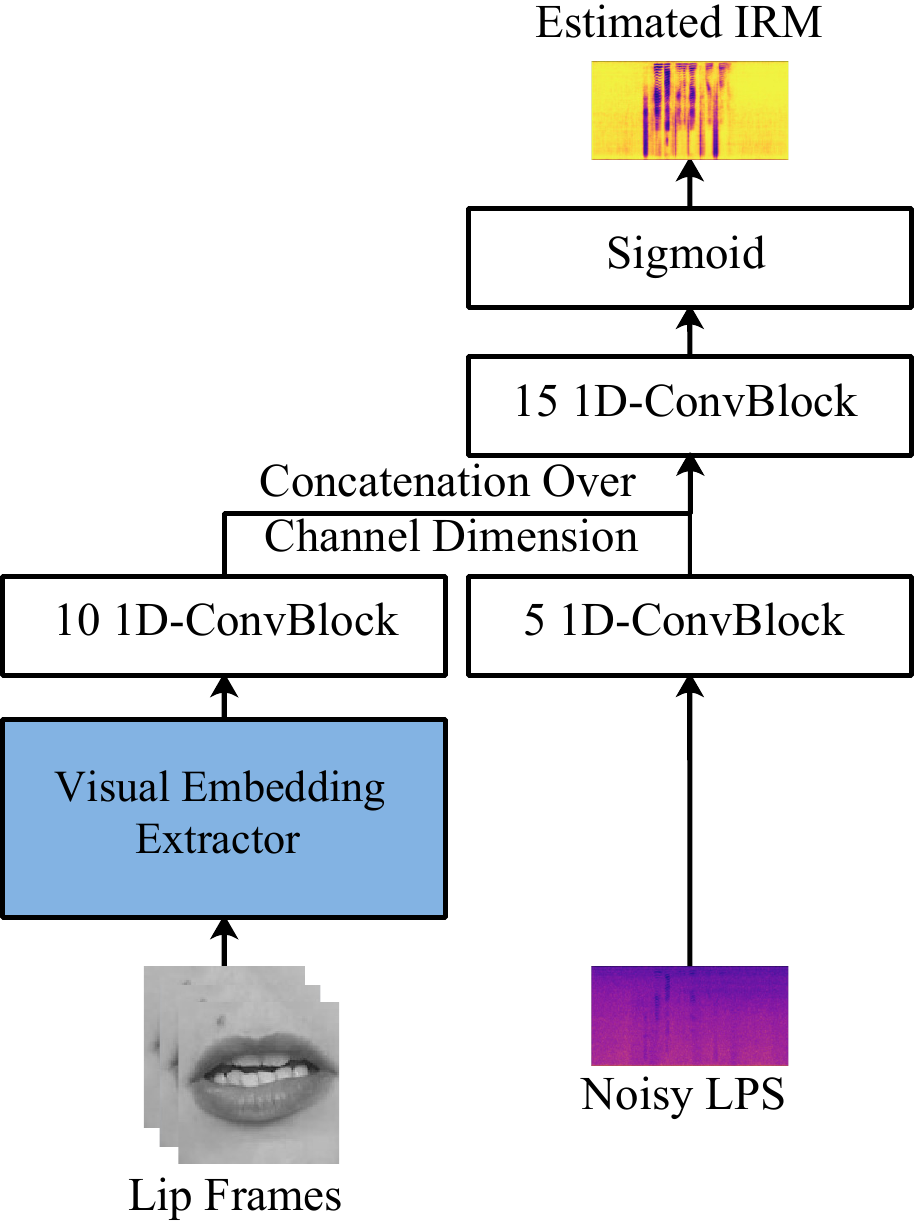}
\caption{Illustration of the VEASE model. The VEASE model takes the visual embeddings as the auxiliary inputs except regular noisy LPS features. The visual embedding extractor is pre-trained separately with classification backend, following the steps introduced in the above-mentioned sections. In the training of the VEASE model, the visual embedding extractor is kept frozen.}
\label{fig: fig_VEASE}
\end{figure}

 The VEASE model consists of three stacks of 1D-ConvBlocks and a frozen visual embedding extractor, as shown in Figure \ref{fig: fig_VEASE}. Each 1D-ConvBlock includes a 1D convolution layer with a residual connection, a ReLU activation, and a batch normalization, as in \cite{Afouras2018}. Some of the blocks contain an extra up-sampling or down-sampling layer, because the number of audio frames is different from that of the video frames.

 Visual embedding $E_{\rm V}$ is processed by the stack $s_{\rm E}$ at the bottom left consisting of $N_{\rm E}$ 1D-ConvBlocks while noisy log-power spectra (LPS) features $A_{\rm LPS} = \{A^t_{\rm LPS}\in\mathbb{R}^{F};t=0,1,\dots,T_{\rm A}-1\}$ are processed by the stack $s_{\rm LPS}$ at the bottom right consisting of $N_{\rm LPS}$ 1D-ConvBlocks:
\begin{align}
R_{\rm E} &= s_{\rm E}(E_{\rm V}) = \overbrace{\operatorname{ConvBlock_{1D}}(\cdots \operatorname{ConvBlock_{1D}}(E_{\rm V}))}^{N_{\rm E}} \\
R_{\rm LPS} &= s_{\rm LPS}(A_{\rm LPS})= \overbrace{\operatorname{ConvBlock_{1D}}(\cdots \operatorname{ConvBlock_{1D}}(A_{\rm LPS}))}^{N_{\rm LPS}}
\end{align}
where $T_{\rm A}$ and $F$ denote the number of time frames and frequency bins for spectrogram, respectively. $R_{\rm E}$ and $R_{\rm LPS}$ denote outputs of different stacks.

 The $R_{\rm E}$ and $R_{\rm LPS}$ are then concatenated along the channel dimension and fed to the top stack $s_{\rm F}$ consisting of $N_{\rm F}$ 1D-ConvBlocks. The last convolution layer in the top stack projects the output's dimension into the same one of noisy magnitude spectrogram. Then, the hidden representation is activated by a sigmoid activation to obtain a magnitude mask $M \in \mathbb{R}^{T_{\rm A}\times F}$:
\begin{equation}
\begin{split}
M &= \sigma(s_{\rm F}([R_{\rm E}, R_{\rm LPS}])) \\ &= \sigma(\overbrace{\operatorname{ConvBlock_{1D}}(\cdots \operatorname{ConvBlock_{1D}}([R_{\rm E}, R_{\rm LPS}]))}^{N_{\rm F}})
\end{split}
\end{equation}
The values of $M$ range from 0 to 1. In this study, we use $N_{\rm E}=10$, $N_{\rm LPS}=5$ and $N_{\rm F}=15$ by default.

 To show the effectiveness of embedding on enhancement performance, we also design a competitive no-embedding version of the EASE model which is stripped of the stack $s_{\rm E}$ at the bottom left and the frozen visual embedding extractor, denoted as NoEASE model. The NoEASE model computes $M$ only using the noisy LPS features as inputs:
\begin{align}
M=\sigma(s_{\rm F}(s_{\rm LPS}(A_{\rm LPS})))
\end{align}

 The ideal ratio mask (IRM) \cite{hummersone2014ideal} is employed as the learning target, which is widely used in monaural speech enhancement \cite{wang2014training}. IRM $M_{\rm IRM} \in \mathbb{R}^{T_{\rm A} \times F}$ is calculated as follows:
\begin{equation}
M_{\rm IRM}=\left(\frac{C_{\rm PS}}{C_{\rm PS}+D_{\rm PS}}\right)^{\frac{1}{2}}
\end{equation}
where $C_{\rm PS}\in \mathbb{R}^{T_{\rm A} \times F}$ and $D_{\rm PS}\in \mathbb{R}^{T_{\rm A} \times F}$ denote power spectrograms of clean speech and noise, respectively.

 The mean square error (MSE) $\mathcal{L}_{\rm MSE}$ between $M$ and $M_{\rm IRM}$ is calculated as the loss function:
\begin{equation}
\mathcal{L}_{\rm MSE} = \operatorname{MSE}(M, M_{\rm IRM}) = \sum \|M-M_{\rm IRM}\|^2_2
\label{eq: eq_mse}
\end{equation}

 We use Adam optimizer to train for $100$ epochs with early stopping when there is no improvement on the validation loss for $10$ epochs. The batch size is $96$. Initial learning rate is set to $0.0001$, which is found by ``LR range test" proposed in \cite{smith2017cyclical}, and halved during training if there is no improvement for $3$ epochs on the validation loss. The best model is selected by the lowest validation loss.

\section{Proposed MEASE Model}\label{sec: sec_MEASE_main}

 In this section, we elaborate our proposed MEASE model. The MEASE model takes the fused audio-visual embedding as the auxiliary input instead of the visual embedding. As described in Section \ref{sec: sec_introduction}, the MEASE model utilizing a complementarity of audio and visual features in an information intersection manner. In order to verify the complementarity between audio and visual embeddings, we design an EASE model that utilizes the audio embedding, denoted as AEASE model. The AEASE model has a similar structure to the VEASE model with the main difference of employing an audio embedding extractor instead of the visual embedding extractor. For verifying the effectiveness of the information intersection-based audio-visual fusion manner on enhancement performance, we design an EASE model that utilizes the concatenation of audio and visual embeddings, denoted as cMEASE model. The details of the AEASE model, the MEASE model and the cMEASE model are elaborated in the following.

\subsection{AEASE model}\label{sec: sec_AEASE_main}

\begin{figure}[htb]
\centering
\includegraphics[width=\linewidth]{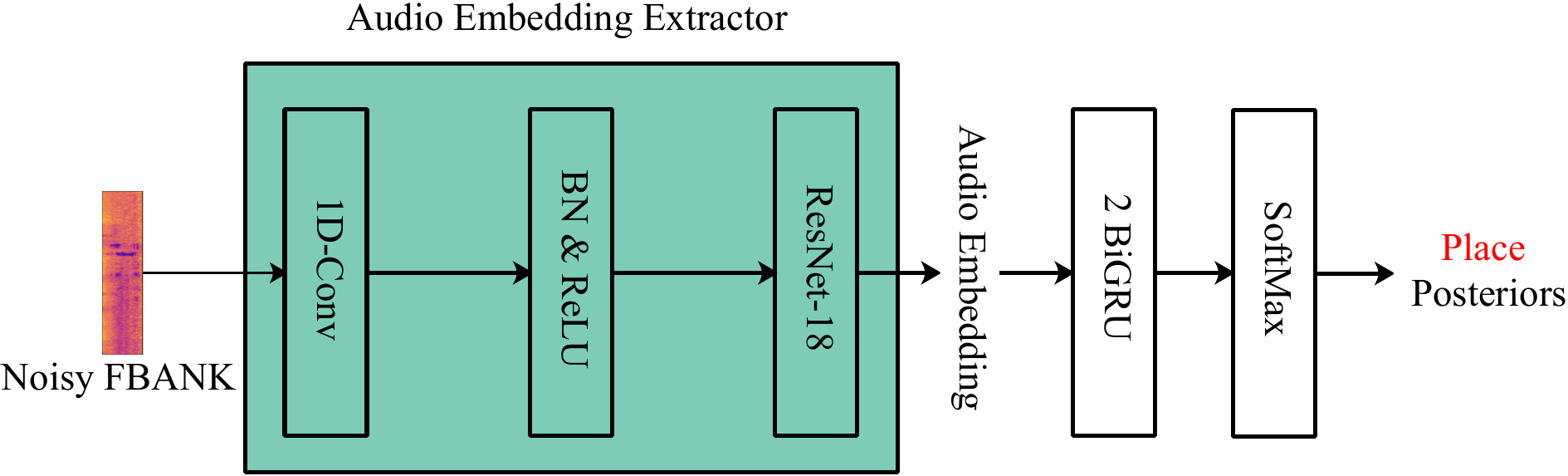}
\caption{Illustration of a audio embedding extractor (in color green for ease of cross-referencing in Figure \ref{fig: fig_AEE}, Figure \ref{fig: fig_MEASE} and Figure \ref{fig: fig_fuse_stage}). The audio embedding extractor has the similar structure as the visual embedding extractor in Section \ref{sec: sec_VEE_arc}. The training of the audio embedding extractor is same as that in Section \ref{sec: sec_VEE_place}.}
\label{fig: fig_AEE}
\end{figure}

 The AEASE model has a similar structure to the VEASE model as shown in Figure \ref{fig: fig_VEASE}, with the main difference of employing a audio embedding extractor, instead of the visual embedding extractor.

 The audio embedding extractor $f_{\rm A}$ has a similar structure as the visual embedding extractor in Section \ref{sec: sec_VEE_arc}, as shown in Figure \ref{fig: fig_AEE}. The 3D-kernels in the spatiotemporal convolution are replaced by 1D-kernels meanwhile the 3D-MaxPooling layer is dropped in this case as the audio frame is a vector. We also use the ResNet-18 with the main difference of employing 1D-kernels instead of 2D-kernels. Given noisy FBANK features $A_{\rm FBANK}\in\mathbb{R}^{T_{\rm A}\times F_{\rm mel}}$, the audio embeddings $E_{\rm A}\in\mathbb{R}^{T_{\rm A}\times L_{\rm A}}$ are calculated as follows:
\begin{equation}
\begin{split}
E_{\rm A} &= \{E^{t}_{\rm A}\in\mathbb{R}^{L_{\rm A}}; t=0, 1, \dots, T_{\rm A}-1\}= f_{\rm A}(A_{\rm FBANK}) \\ &= \operatorname{ResNet-18_A}(\operatorname{BN}(\operatorname{ReLU}(\operatorname{Conv_{1D}}(A_{\rm FBANK}))))
\end{split}
\end{equation}
where, $F_{\rm mel}$ and $L_{\rm A}$ are the number of triangular filters set for FBANK features and the length of $E^{t}_{\rm A}$, respectively. In this study, $L_{\rm A}=L_{\rm V}=256$ is used by default.

 We use the same training process to train the audio embedding extractor as training the visual embedding extractor in Section \ref{sec: sec_VEE_place}. Adam optimizer is used to minimize $\mathcal{L}_{\rm CE}$, which is calculated by Equation (\ref{eq: eq_place_ce}). But $P_{\rm place}$ is computed by using $E_{\rm A}$:
\begin{equation}
P_{\rm place} = f_C(E_{\rm A})
\end{equation}

 The AEASE model takes both $A_{\rm LPS}$ and $E_{\rm A}$ as inputs and outputs $M$:
\begin{align}
M=\sigma(s_{\rm F}([s_{\rm E}(E_{\rm A}), s_{\rm LPS}(A_{\rm LPS})]))
\end{align}
The same optimization process as in Section \ref{sec: sec_VEASE} is also used to minimize $\mathcal{L}_{\rm MSE}$, which is calculated by Equation (\ref{eq: eq_mse}).

\subsection{MEASE Model}\label{sec: sec_MEASE}

\begin{figure}[htb]
\centering
\includegraphics[width=0.8\linewidth]{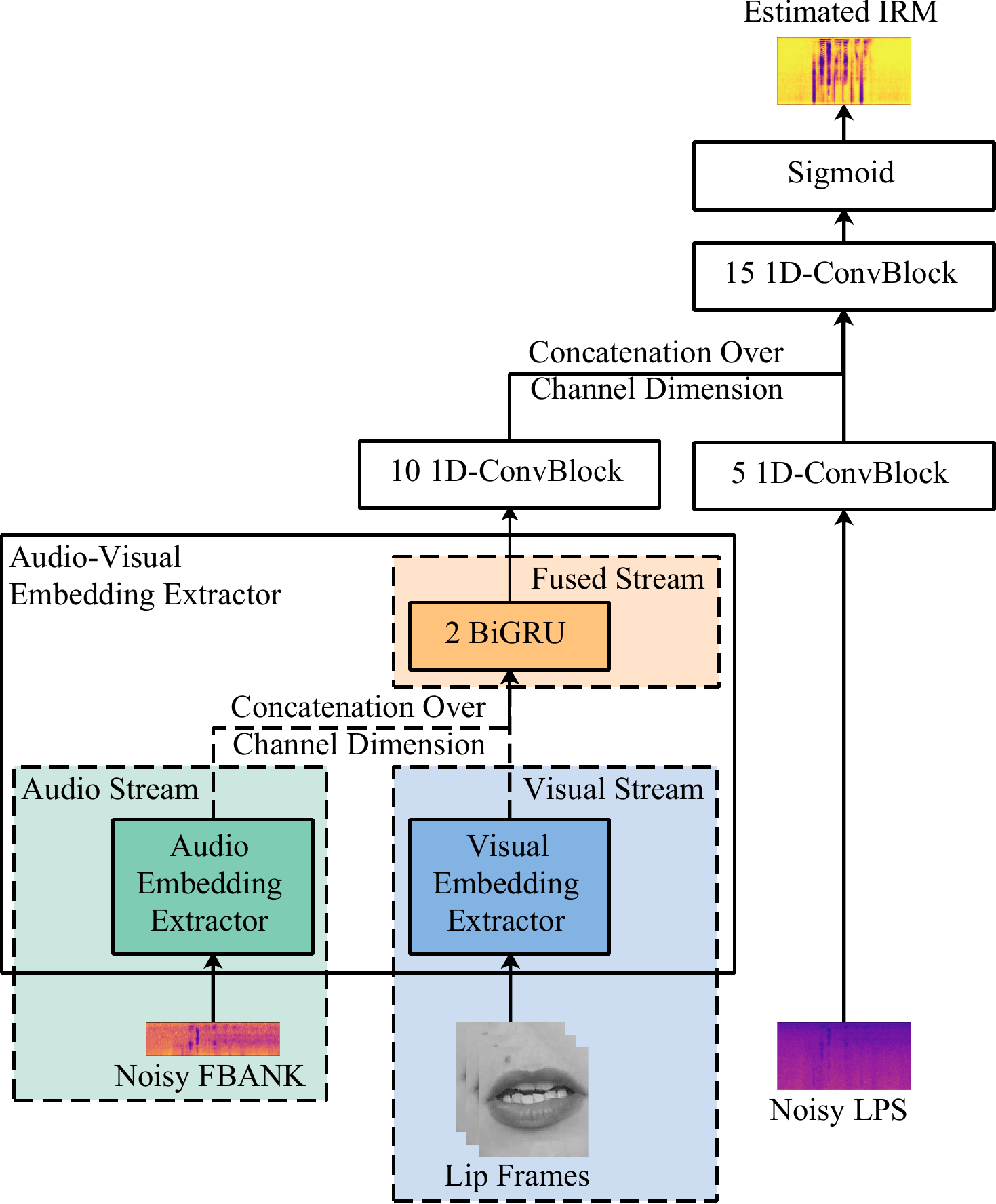}
\caption{Illustration of the proposed MEASE model. The pervious visual embedding extractor evolves to the audio-visual embedding extractor, which consists of a visual stream (in blue), an audio stream (in green) and a fused stream (in orange). The audio-visual embedding extractor fuse the audio and visual embeddings in an information intersection manner.}
\label{fig: fig_MEASE}
\end{figure}

 The most significant change in the MEASE model is that the visual embedding extractor evolves into the audio-visual embedding extractor. The audio-visual embedding extractor takes not only lip frames but also noisy FBANK features as inputs and outputs the fused audio-visual embedding which is learned under the supervision of the information intersection, i.e., the articulation place label.

 The audio-visual embedding extractor consists of visual, audio and fused streams, as shown at the bottom left part of Figure \ref{fig: fig_MEASE}. The visual stream has the same structure as the visual embedding extractor in Section \ref{sec: sec_VEE_arc} while the audio stream has the same structure as the audio embedding extractor in Section \ref{sec: sec_AEASE_main}. $V$ and $A_{\rm FBANK}$ are processed by visual and audio streams, respectively:
\begin{align}
E^{\rm V}_{\rm AV} &= f_{\rm V}(V) \\
E^{\rm A}_{\rm AV} &= f_{\rm A}(A_{\rm FBANK})
\end{align}
where $E^{\rm V}_{\rm AV}\in\mathbb{R}^{T_{\rm V}\times L_{\rm V}}$ and $E^{\rm A}_{\rm AV}\in\mathbb{R}^{T_{\rm A}\times L_{\rm A}}$ denotes the outputs of visual and audio streams, respectively. The mismatch in the number of frames between $E^{\rm V}_{\rm AV}$ and $E^{\rm A}_{\rm AV}$, i.e., $T_{\rm A} \ne T_{\rm V}$, is solved by repeating a video frame for several audio frames:
\begin{equation}
\tilde{E}^{\rm V}_{\rm AV} = \{\overbrace{E^{{\rm V}, 0}_{\rm AV}, \cdots, E^{{\rm V}, 0}_{\rm AV}}^{T_{\rm A}/T_{\rm V}}, E^{{\rm V}, 1}_{\rm AV}\cdots\}
\end{equation}

 The fused stream consisting of a $2$-layers BiGRU at the top takes $\tilde{E}^{\rm V}_{\rm AV}$ and $E^{\rm A}_{\rm AV}$ as inputs and outputs the audio-visual embedding $E_{\rm AV}\in\mathbb{R}^{T_{\rm A}\times L_{\rm AV}}$:
\begin{equation}
E_{\rm AV} = \{E^{t}_{\rm AV}; t=0, 1, \cdots, T_{\rm A}-1\} = \operatorname{BiGRU}(\operatorname{BiGRU}([\tilde{E}^{\rm V}_{\rm AV}, E^{\rm A}_{\rm AV}]))
\end{equation}
where $L_{\rm AV}$ is the length of $E^t_{\rm AV}$. In this paper, we use $L_{\rm AV}=L_{\rm A}+L_{\rm V}=512$ by default.

 We also use the same steps to minimize $\mathcal{L}_{\rm CE}$, which is calculated by Equation (\ref{eq: eq_place_ce}), as these in Section \ref{sec: sec_VEE_place}. But $P_{\rm place}$ is computed by using $E_{\rm AV}$:
\begin{equation}
P_{\rm place} = f_C(E_{\rm AV})
\end{equation}
It is a remarkable fact that the audio-visual classification model can achieve a better and faster convergence, by initializing visual and audio streams with the independently pre-trained params.

 The MEASE model takes both $A_{\rm LPS}$ and $E_{\rm AV}$ as inputs and outputs $M$:
\begin{align}
M=\sigma(s_{\rm F}([s_{\rm E}(E_{\rm AV}), s_{\rm LPS}(A_{\rm LPS})]))
\end{align}
We use the same optimization process as in Section \ref{sec: sec_VEASE} to minimize $\mathcal{L}_{\rm MSE}$, which is calculated by Equation (\ref{eq: eq_mse}).

\subsection{cMEASE Model} \label{sec: sec_cMEASE}

 By ablating the fused stream in Figure \ref{fig: fig_MEASE}, another audio-visual embedding, $cE_{\rm AV}\in\mathbb{R}^{T_{\rm A}\times (L_{\rm A}+L_{\rm V})}$, which is the concatenation of audio and visual embeddings, is designed:
\begin{equation}
cE_{\rm AV} = [E_{\rm V}, E_{\rm A}] = [f_{\rm V}(V), f_{\rm A}(A_{\rm FBANK})]
\label{eq: eq_ceav}
\end{equation}
where $f_{\rm A}$ and $f_{\rm V}$ are trained independently, following the steps introduced in Section \ref{sec: sec_AEASE_main} and Section \ref{sec: sec_VEE_place}, respectively.

The cMEASE model takes both $A_{\rm LPS}$ and $cE_{\rm AV}$ as inputs and outputs $M$:
\begin{align}
M=\sigma(s_{\rm F}([s_{\rm E}(cE_{\rm AV}), s_{\rm LPS}(A_{\rm LPS})]))
\end{align}
We use the same optimization process as in Section \ref{sec: sec_VEASE} to minimize $\mathcal{L}_{\rm MSE}$, which is calculated by Equation (\ref{eq: eq_mse}).

\subsection{Fusion Stage of Audio and Visual Embeddings}\label{sec: sec_fuse_stage}

\begin{figure}[htb]
\centering
\includegraphics[width=0.95\linewidth]{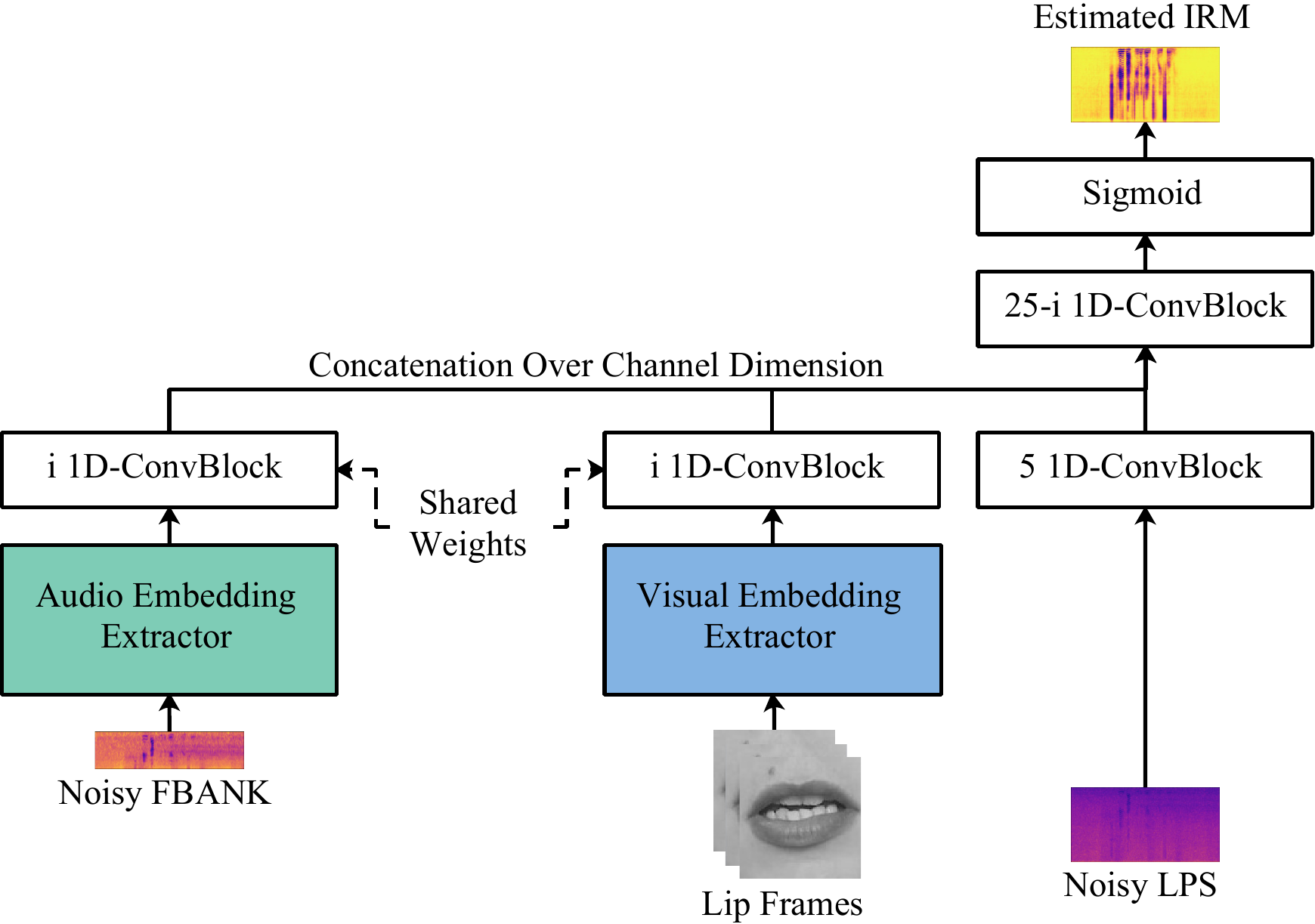}
\caption{Illustration of the MEASE model with different fusion stages of audio and visual embeddings.}
\label{fig: fig_fuse_stage}
\end{figure}

 To study the effect of the audio-visual fusion stage on enhancement performance, we design a MEASE model that fuses visual and audio embeddings at the $i$-th layer of the enhancement network, denoted as Middle-i model, as shown in Figure \ref{fig: fig_fuse_stage}. We change $N_{\rm E}$ with the fixed sum of $N_{\rm E}$ and $N_{\rm F}$ and use the same stack to process audio and visual embeddings, respectively:
\begin{align}
R_{E_{\rm V}}=& s_{\rm E}(E_{\rm V}) = \overbrace{\operatorname{ConvBlock_{1D}}(\cdots \operatorname{ConvBlock_{1D}}(E_{\rm V}))}^{N_{\rm E} = i} \label{eq: eq_VE} \\
R_{E_{\rm A}} =& s_{\rm E}(E_{\rm A}) = \overbrace{\operatorname{ConvBlock_{1D}}(\cdots \operatorname{ConvBlock_{1D}}(E_{\rm A}))}^{N_{\rm E} = i} \label{eq: eq_AE} \\
R_{\rm LPS}=& s_{\rm LPS}(A_{\rm LPS}) = \overbrace{\operatorname{ConvBlock_{1D}}(\cdots \operatorname{ConvBlock_{1D}}(A_{\rm LPS}))}^{N_{\rm LPS}}\\
\begin{split}
M=&\sigma(s_{\rm F}([R_{E_{\rm V}}, R_{E_{\rm A}}, R_{\rm LPS}])) \\
=&\sigma(\overbrace{\operatorname{ConvBlock_{1D}}(\cdots \operatorname{ConvBlock_{1D}}([R_{E_{\rm V}}, R_{E_{\rm A}}, R_{\rm LPS}]))}^{N_{\rm F} = 25-i})
\end{split}
\end{align}
where $s_{\rm E}(\cdot)$ in Equation (\ref{eq: eq_VE}) has the same params as that in Equation (\ref{eq: eq_AE}), as well as $E_{\rm A}$ and $E_{\rm V}$ are extracted by using $f_{\rm A}$ and $f_{\rm V}$ trained independently. By modifying the value of $i$, we can make the fusion take places at different stages without changing the network structure.

\section{Experiments}\label{sec: sec_exp}

\subsection{Dataset}\label{sec: sec_dataset}

 To evaluate the performance of our proposed method, we created a simulation dataset of noisy speech based on the TCD-TIMIT audio-visual corpus \cite{harte2015tcd}. The TCD-TIMIT consisted of $59$ volunteer speakers with around $98$ videos each, as well as $3$ lipspeakers who specially were trained to speak in a way that helped the deaf understand their visual speech. The speakers were recorded saying various sentences from the TIMIT corpus \cite{garofolo1993darpa} by using both front-facing and $30$-degree cameras. However, the utterances of $3$ lipspeakers and $30$-degree videos were not used in this paper. For testing the robustness to unseen speaker condition, we divided these videos and audios into a \emph{train-clean} set which consisted of $57$ speakers ($31$ male and $26$ female) and a \emph{test-clean} set which consisted of $2$ speakers ($1$ male and $1$ female) who were not in the \emph{train-clean} set.

We chose the TCD-TIMIT dataset for two main reasons:

\begin{enumerate}[(1)]

\item TCD-TIMIT was recorded in a controlled environment, and provided near-field signals collected by a microphone close to the mouth, which can ensure that the utterances do not contain background noise. While other large-scale in-the-wild audio-visual datasets, such as BBC-Oxford LipReading Sentences 2 (LRS2) dataset \cite{chung2017lip}, AVSpeech dataset \cite{Ephrat2018looking}, etc, were collected from real-world sources using automated pipeline, and none of them was checked whether background noise exists.\footnote{We manually listen to the test and verification sets of the LRS2 dataset. We find more than half of sentences can be clearly perceived as noisy.} When testing an enhancement system, if the ground truth contains background noise, the metrics will be severely distorted and cannot well measure the performance of the system.

\item The utterances consisted of various phrases in the TCD-TIMIT dataset, thus they were more suitable for actual scenarios than the utterances consisting of a fixed set of phrases in the GRID dataset \cite{cooke2006audio}. The TCD-TIMIT dataset also contained phonetic-level transcriptions, which provided available labels for the embedding extractor training.
\end{enumerate}

 A total of $115$ noise types, including $100$ noise types in \cite{hu2010tandem} and $15$ homemade noise types, were adopted for training to improve the robustness to unseen noise types. The $5600$ utterances from \emph{train-clean} set were corrupted with the above-mentioned $115$ noise types at five levels of SNRs, i.e., $15$ dB, $10$ dB, $5$ dB, $0$ dB and $-5$ dB, to build an $35$-hour multi-condition training set consisting of pairs of clean and noisy utterances. The other $43$ utterances from \emph{train-clean} set were corrupted with $3$ unseen noise types at above-mentioned SNR levels to build a validation set, i.e., Destroyer Operations, Factory2 and F-16 Cockpit. The $198$ utterances from \emph{test-clean} set were used to construct a test set for each combination of $3$ other unseen noise type and above SNR levels, i.e., Destroyer Engine, Factory1 and Speech Babble. All unseen noise were collected from the NOISEX-92 corpus \cite{varga1993assessment}. The five levels of SNRs in the training set were also adopted for testing and validating.

 For audio preprocessing, all speech signals were resampled to $16$ kHz. A $400$-point short-time Fourier transform was used to compute the spectra of each overlapping windowed frame. Here, a $25$-ms Hanning window and a $10$-ms window shift were adopted. In our experiments, $201$-dimensional LPS vectors were generated to train the enhancement network and $40$-dimensional FBANK vectors were generated to train the embedding extractor, i.e., $F=201, F_{\rm mel}=40$. Mean and variance normalizations were applied to the noisy LPS and FBANK vectors.

 For video preprocessing, a given video clip was downsampled from $29.97$ fps to $25$ fps, i.e., $T_A=4\times T_V$. For every video frame, $68$ facial landmarks were extracted by using Dlib \cite{king2009dlib} implementation of the face landmark estimator described in \cite{kazemi2014one}, then we cropped a lip-centered window of size $98\times 98$ pixels by using the $20$ lip landmarks from the $68$ facial landmarks. The frames were transformed to grayscale and normalized with respect to the overall mean and variance.

\subsection{Evaluation Protocol}\label{sec: sec_metric}

 In this experiment, we mainly adopt Perceptual Evaluation of Speech Quality (PESQ) \cite{rix2001perceptual} and Short-Time Objective Intelligibility (STOI) \cite{taal2011algorithm} to evaluate models. Both metrics are commonly used to evaluate the performance of speech enhancement system. PESQ, which is a speech quality estimator, is designed to predict the mean opinion score of a speech quality listening test for certain degradations. Moreover, to show the improvement in speech intelligibility, we also calculated STOI. The STOI score is typically between $0$ and $1$, and the PESQ score is between $-0.5$ and $4.5$. For both metrics, higher scores indicate better performance.

\subsection{Results of VEASE Models Utilizing Different Visual Embeddings}\label{sec: sec_best_VE}
\begin{figure}[htb]
\centering
\includegraphics[width=0.8\linewidth]{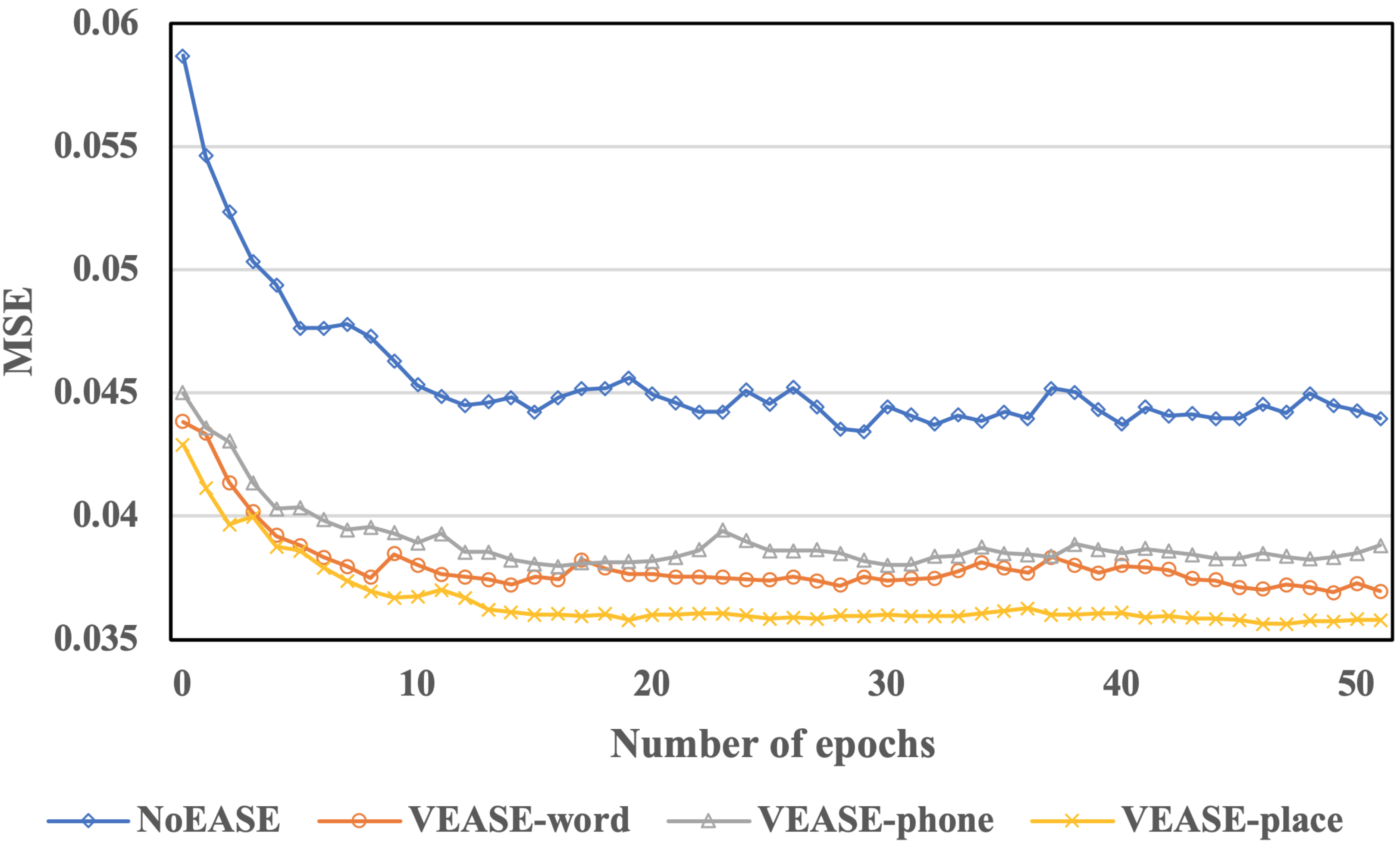}
\caption{A comparison of learning curves among NoEASE, VEASE-word (LRW), VEASE-phone and VEASE-place on the validation set.}
\label{fig: fig_VEASE_val_loss}
\end{figure}

\begin{table}[htb]
\centering
\caption{Average performance comparison of VEASE models with different visual embeddings on the test set at different SNRs averaged over 3 unseen noise types.}
\label{tab: tab_VEASE}
\footnotesize
\setlength{\tabcolsep}{5pt}{
\begin{tabular}[b]{|l|ccccc|ccccc|}
\toprule
\textbf{Model}&\multicolumn{5}{c|}{\textbf{PESQ}}&\multicolumn{5}{c|}{\textbf{STOI}(in \%)}\\
\midrule
{\textbf{SNR}(in dB)}&-5&0&5&10&15&-5&0&5&10&15\\
\midrule
{Noisy}&$1.70$&1.97&2.26&2.56&2.86&54.34&65.11&75.33&84.48&90.88\\
{NoEASE}&2.07&2.34&2.64&2.92&3.21&58.79&70.29&80.24&87.83&92.57\\
{VEASE-word}&2.16&2.45&2.72&2.99&3.25&66.26&75.11&82.57&88.75&92.98\\
{VEASE-phone}&2.14&2.42&2.69&2.96&3.23&66.29&74.89&82.22&88.45&92.79\\
{VEASE-place}&2.21&2.47&2.73&3.00&3.26&66.57&75.27&82.64&88.80&92.96\\
\bottomrule
\end{tabular}
}
\end{table}

 In Section \ref{sec: sec_VEASE_main}, we proposed two VEASE models with different visual embeddings, i.e., VEASE-phone and VEASE-place. To compare their effectiveness with the baseline model, i.e., VEASE-word (LRW), on enhancement performance, a series of experiments were conducted for the unprocessed system denoted as Noisy, NoEASE, VEASE-word (LRW), VEASE-phone and VEASE-place. We present the learning curves of the MSEs among NoEASE, VEASE-word (LRW) and VEASE-phone and VEASE-place on the validation set in Figure \ref{fig: fig_VEASE_val_loss}. The corresponding evaluation metrics are shown in Table \ref{tab: tab_VEASE}. We evaluate the average performance of two measures at different SNRs across $3$ unseen noise types.

 Based on Figure \ref{fig: fig_VEASE_val_loss} and Table \ref{tab: tab_VEASE}, we find following observations.

\begin{enumerate}[(1)]
\item The learning curves of the MSEs indicate that all VEASE models consistently generate smaller MSEs on the validation set than NoEASE. This result implies that the visual embedding is useful for speech enhancement. As shown in Table \ref{tab: tab_VEASE}, all VEASE models yield improvements in PESQ and STOI over NoEASE in all SNRs. In particular, the improvement is more significant at low SNRs cases.
\item VEASE-word (LRW) consistently yields smaller MSE values and a slower convergence than VEASE-phone. This implies that the visual embedding in VEASE-word (LRW) is more useful but slow-fit for speech enhancement. In VEASE-word (LRW), visual embedding extractor is trained with a large amount of additional data, so it can obtain better generalization ability and provide visual embedding with more information. On the other side, the data mismatch between embedding learning and enhancement task brings information redundancy, which leads to a slower convergence. This observation is consistent with the comparison of the objective evaluation metrics on the test set shown in Table \ref{tab: tab_VEASE}. And VEASE-phone does not perform better than VEASE-word (LRW) in any evaluation metrics.
\item VEASE-place clearly achieves a better and faster convergence than VEASE-phone and VEASE-word (LRW). This implies that VEASE-place provides more useful and quick-fit visual embedding for speech enhancement. By comparing the evaluation metrics in Table \ref{tab: tab_VEASE}, we also observe that VEASE-place not only yields remarkable gains over VEASE-phone across all evaluation metrics and all SNR levels, but also outperforms VEASE-word (LRW) in most cases with only one exception for the STOI at $15$ dB SNR. And in that exceptional situation, the results are still close. These results suggest that our proposed VEASE-place model achieves a better generalization capability, while reducing mismatch between embedding learning and enhancement task.
\end{enumerate}

 Overall, the high correlation between the articulation place label and the acoustic information in video is beneficial to the extraction of visual embedding, which is useful for speech enhancement, even if no requirement of additional data. Therefore, we select articulation place as the default classification target in all subsequent experiments and use VEASE to refer to VEASE-place in all subsequent sections.

\subsection{Results of Proposed MEASE Model} \label{sec: sec_MEASE_result}

\begin{table}[htb]
\centering
\caption{Average performance comparison of NoEASE model, VEASE model, AEASE model, cMEASE model and MEASE model on the test set at different SNRs averaged over 3 unseen noise types.}
\label{tab: tab_MEASE_result}
\footnotesize
\setlength{\tabcolsep}{5pt}{
\begin{tabular}[b]{|l|ccccc|ccccc|}
\toprule
\textbf{Model}&\multicolumn{5}{c|}{\textbf{PESQ}}&\multicolumn{5}{c|}{\textbf{STOI}(in \%)}\\
\midrule
 \textbf{SNR}(in dB)&-5&0&5&10&15&-5&0&5&10&15\\
\midrule
NoEASE&2.07&2.34& 2.64&2.92&3.21&58.79&70.29&80.24&87.83&92.57\\
VEASE&2.21&2.47&2.73&3.00&3.26&66.57&75.27&82.64&88.80&92.96\\
AEASE&2.09&2.39&2.69&2.98&3.27&60.84&72.24&81.58&88.39&92.76\\
cMEASE&2.27&2.55&2.81&3.08&3.34&67.60&76.26&83.26&89.13&93.12\\
MEASE&2.29&2.59&2.88&3.16&3.42&68.96&77.64&84.43&89.99&93.64\\
\bottomrule
\end{tabular}
}
\end{table}

 In this section, the goal is to examine the effectiveness of the proposed MEASE model on enhancement performance, and obtain a better understanding about the contribution of different parts of the MESAE model. We present an average performance comparison between NoEASE, VEASE, AEASE, cMEASE and MEASE in Table \ref{tab: tab_MEASE_result}.

 Paying attention to the last row in Table \ref{tab: tab_MEASE_result}, we can observe that MEASE shows significant improvements over VEASE across all evaluation metrics, and larger gains are observed at high SNRs. By comparing the results of VEASE with NoEASE, the improvement yielded by visual embedding decreases as SNR increases, for example, the PESQ of VEASE increased from $2.07$ to $2.21$ at $-5$ dB SNR and from $3.21$ to $3.26$ at $15$ dB SNR. This observation is consistent with that in \cite{wang2020robust}. In contrast, MEASE shows stable improvements over NoEASE for high SNRs. For example, the PESQ of MEASE increased from $2.07$ to $2.29$ at $-5$ dB SNR and from $3.21$ to $3.42$ at $15$ dB SNR. All these results indicate that MEASE is more robust against the change of noise level and yields better generalization capability than VEASE.

 Table \ref{tab: tab_MEASE_result} also shows the results of AEASE. By comparing its results with NoEASE, we can observe that the improvement yielded by audio embedding increases as SNR grows, for example, the PESQ of AEASE increased from $2.07$ to $2.09$ at $-5$ dB SNR and from $3.21$ to $3.27$ at $15$ dB SNR. This suggest that the complementarity between audio and visual embeddings lies in the variation tendencies of metric improvement with respect to SNR level. But directly comparing AEASE and VEASE on the evaluation metrics as shown in Table \ref{tab: tab_MEASE_result}, we can not observe that AEASE performs better than VEASE at high SNRs, i.e., $\operatorname{SNR}=5$, $10$ and $15$ dB, especially at $5$ dB SNR.

\begin{table}[htb]
\centering
\caption{Average performances of different models on the test set at different SNRs and different articulation places averaged over 3 unseen noise types.}
\label{tab: tab_place_pesq}
\tiny
\setlength{\tabcolsep}{0.6pt}{
\begin{tabular}[b]{|l|c|c|c|c|c|c|c|c|c|c|c|c|}
\toprule
\textbf{SNR}(in dB)&\multicolumn{4}{c|}{-5}&\multicolumn{4}{c|}{0}&\multicolumn{4}{c|}{5}\\
\midrule
\diagbox{Place}{Model}&NoEASE&AEASE&VEASE&MEASE&NoEASE&AEASE&VEASE&MEASE&NoEASE&AEASE&VEASE&MEASE\\
\midrule
Labial&1.28&1.38&1.58&1.76&1.57&1.75&1.81&2.06&2.05&2.18&2.23&2.50\\
Mid&1.54&1.68&1.86&2.02&2.03&2.21&2.29&2.45&2.58&2.72&2.73&2.96\\
High&1.38&1.52&1.65&1.81&1.79&1.95&1.99&2.17&2.28&2.39&2.42&2.62\\
Low&1.63&1.89&2.00&2.29&2.17&2.48&2.46&2.69&2.84&2.99&2.93&3.20\\
Retroflex&1.46&1.66&1.75&2.00&1.95&2.15&2.12&2.32&2.44&2.57&2.54&2.77\\
Coronal&1.59&1.74&1.80&1.93&1.92&2.07&2.05&2.23&2.30&2.39&2.35&2.56\\
Glottal&1.02&1.22&1.36&1.70&1.42&1.71&1.59&1.92&1.95&2.10&2.05&2.30\\
Velar&1.31&1.44&1.41&1.49&1.48&1.64&1.68&1.86&1.86&2.01&2.00&2.22\\
Dental&0.94&1.22&1.25&1.64&1.32&1.62&1.36&2.05&1.98&2.21&1.98&2.44\\
\bottomrule
\end{tabular}
}
\end{table}

 To further explore the complementarity between audio and visual embeddings, we present an average performance comparison between utterance segments belonging to different articulation places in Table \ref{tab: tab_place_pesq}. Because the utterance segment does not have actual semantics, we only examine the average performance of PESQ at different SNRs across $3$ unseen noise types. Table \ref{tab: tab_place_pesq} illustrates VEASE and AEASE play a major role at different articulation places, respectively, at the same SNR level. Even at high SNRs, VEASE still yields improvement than AEASE in some articulation places. For example, VEASE's PESQ values are $2.23$, $2.73$, $2.42$, while AEASE's PESQ values are $2.18$, $2.72$, $2.39$ in Labial, Mid, High at $5$ dB SNR level. This result explains why AEASE does not outperform VEASE at high SNR levels. Relating to the lip shapes belonging to different articulation places, as shown in Figure \ref{fig: fig_lip2place}, we find VEASE yields greater improvement at articulation places where the lip shapes change greatly, i.e., Labial, Mid and High, while AEASE is on the contrary. Overall, we can conclude that the complementarity between audio and visual embeddings lies in different SNR levels, as well as different articulation places. More specifically, in the cases that the SNR level is low and the articulation place has high visual correlation, visual embedding performs better. And audio embedding is better on articulation places with low visual correlation at high SNR levels. Based on these observations, our proposed MEASE model takes the advantages of visual and audio embeddings, and achieves the best performance in all SNRs and all articulation places.

 The information intersection-based audio-visual fusion manner in the MEASE model is our another contribution. From Table \ref{tab: tab_MEASE_result}, we can observe that MEASE consistently outperforms cMEASE over all SNR levels in terms of all 2 measures, especially at high SNRs. This observation demonstrates that the information intersection-based audio-visual fusion method has better information integration capability for audio and visual embeddings than channel-wise concatenation which is widely used in previous works.

\subsection{Results of Different Audio-Visual Fusion Stages}\label{sec: sec_fusion_posion}

\begin{figure}[htbp]
\centering
\subfigure[PESQ]{
\begin{minipage}[t]{\linewidth}
\centering
\includegraphics[width=0.8\linewidth]{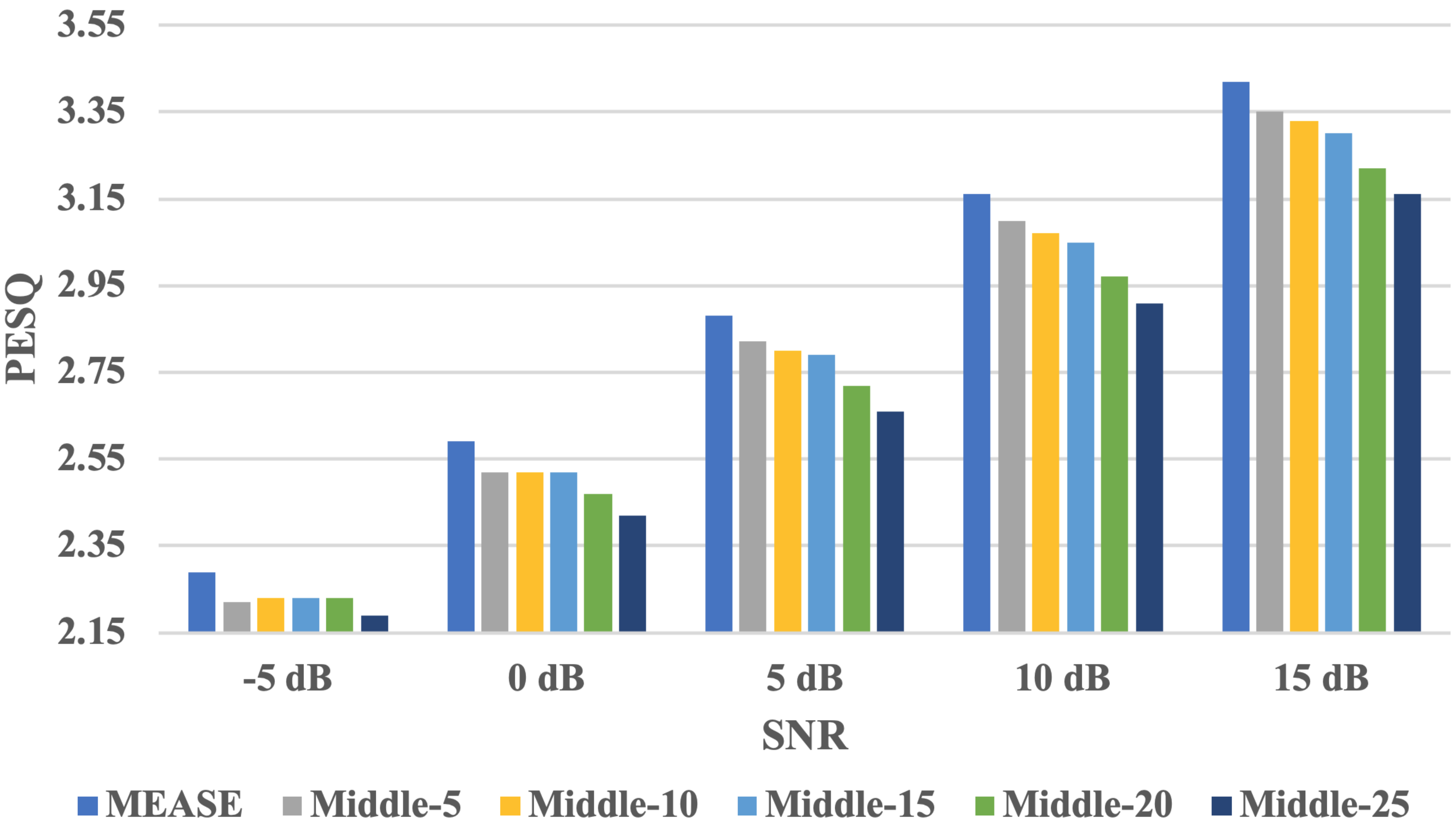}
\label{fig: fig_fusion_stage_result_PESQ}
\end{minipage}%
}%

\subfigure[STOI]{
\begin{minipage}[t]{\linewidth}
\centering
\includegraphics[width=0.8\linewidth]{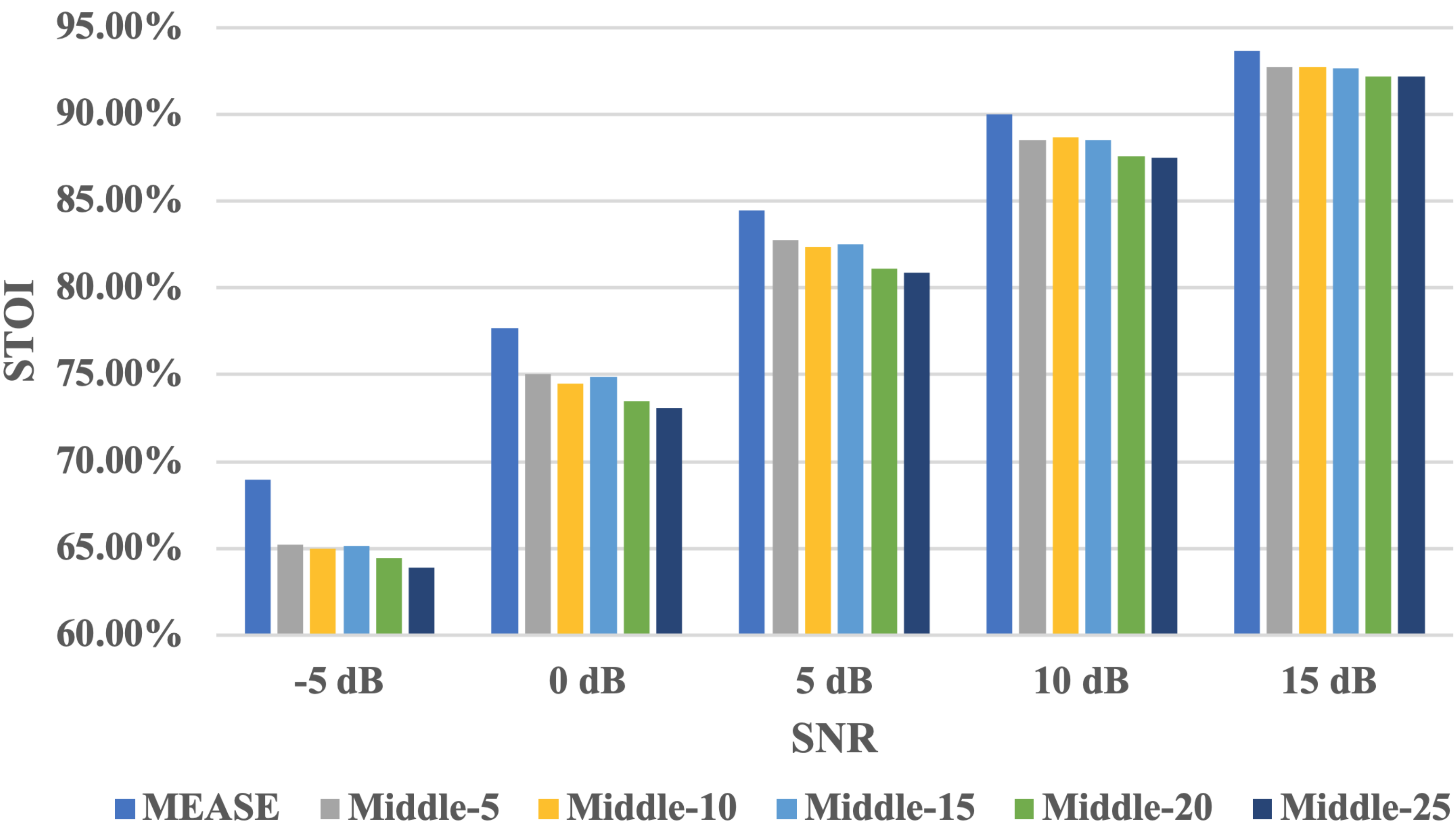}
\label{fig: fig_fusion_stage_result_STOI}
\end{minipage}%
}%
\centering
\caption{Average performance comparison among different audio-visual fusion stages for the PESQ/STOI measures at different SNRs averaged over $3$ unseen noise types. The top figure shows the PESQ measure. The bottom figure shows the STOI measure.}
\label{fig: fig_fusion_position_result}
\end{figure}

 One of the most significant differences between our method and previous methods is that the proposed MEASE model fuses audio and visual modes in the stage of embedding extractor training. It is an early fusion while previous methods fuse audio and visual modes in the middle of the enhancement network, known as the middle fusion. For verifying the effectiveness of the early fusion on enhancement performance, we design an experimental comparative study described in Section \ref{sec: sec_fuse_stage} and conduct a set of experiments using five different $i$, i.e., $i=5, 10, 15, 20, 25$.

 As we can see from Figure \ref{fig: fig_fusion_position_result}, MEASE achieves the best results over all models utilizing the middle fusion across all evaluation metrics for all SNR levels. By comparing the results of different middle fusion-based models, the variation tendencies of all objective metrics with respect to different fusion stage get worse as the stage moves back. These results suggest that early fusion strategy can better integrate useful information for speech enhancement from both modalities than the standard fusion which happens at the middle layer of enhancement network.

\section{Conclusion} \label{sec: sec_conclusion}

 In this study, we extend the previous audio-visual speech enhancement (AVSE) framework to embedding aware speech enhancement (EASE). We first propose visual embedding to enhance speech, leveraging upon the high correlation between articulation place labels and acoustic information in videos. Next, we propose multi-modal audio-visual embedding obtained by fusing audio and visual embedding in the stage of embedding extraction training under the supervision of their information intersection at the articulation place label level.

 Extensive experiments empirically validate that our proposed visual embedding consistently yields improvements over the conventional word-based approaches. And our proposed audio-visual embedding achieves even greater performance improvements by utilizing the complementarity of audio and visual embedding in a information intersection-based way, with higher information integration capabilities and better speech enhancement performance in early fusion.

 Our future work will focus on how to use unsupervised or self-supervised techniques to extract effective audio-visual embedding, in order to achieve a comparable or better enhancement performance than the current framework.

\section{Acknowledgements}
This work was supported by the National Natural Science Foundation of China under Grant No.61671422.


\bibliography{mybibfile}

\end{document}